\def\be{\begin{equation}}
\def\ee{\end{equation}}
\def\ba{\begin{array}}
\def\ea{\end{array}}

\documentclass[aps,amsmath,amssymb,amsfonts,showpacs,twocolumn,pra]{revtex4}
\usepackage{graphicx}
\usepackage{epstopdf}
\def\qed{\leavevmode\unskip\penalty9999 \hbox{}\nobreak\hfill
     \quad\hbox{\leavevmode  \hbox to.77778em{%
               \hfil\vrule   \vbox to.675em%
               {\hrule width.6em\vfil\hrule}\vrule\hfil}}
     \par\vskip3pt}

\newtheorem{theorem}{Theorem}

\begin{document}
\title{ Generalized Monogamy Relations of Concurrence for $N$-qubit Systems}
\author{Xue-Na Zhu$^{1}$}
\author{Shao-Ming Fei$^{2,3}$}

\affiliation{$^1$School of Mathematics and Statistics Science, Ludong University, Yantai 264025, China\\
$^2$School of Mathematical Sciences, Capital Normal University,
Beijing 100048, China\\
$^3$Max-Planck-Institute for Mathematics in the Sciences, 04103
Leipzig, Germany}

\begin{abstract}

We present a new kind of monogamous relations based on concurrence and concurrence of assistance.
For $N$-qubit systems $ABC_1...C_{N-2}$, the monogamy relations satisfied by the concurrence of $N$-qubit pure states
under the partition $AB$ and $C_1...C_{N-2}$, as well as under the partition $ABC_1$ and $C_2...C_{N-2}$
are established, which give rise to a kind of restrictions on the entanglement distribution and trade off
among the subsystems.

\end{abstract}

\maketitle

Quantum entanglement \cite{t1,t2,t3,t4,t5,t6} is an essential feature of quantum mechanics,
which distinguishes the quantum from classical world.
As one of the fundamental differences between quantum entanglement and classical
correlations, a key property of entanglement is that a quantum system entangled with one of other
systems limits its entanglement with the remaining others.
In multipartite quantum systems, there can be several inequivalent types entanglement among
the subsystems and the amount of entanglement with different types might not be directly comparable to each other.
The monogamy relation of entanglement is a way to characterize the different types of entanglement distribution.
The monogamy relations give rise to the structures of entanglement in
the multipartite setting. Monogamy is also an essential feature
allowing for security in quantum key distribution \cite{k3}.
Monogamy relations are not always satisfied by entanglement measures.
Although the concurrence and entanglement of formation do not satisfy
such monogamy inequality, it has been shown that
the $\alpha$th $(\alpha\geq2)$ power
of concurrence and $\alpha$th $(\alpha\geq\sqrt{2})$ power entanglement of formation for $N$-qubit states
do satisfy the monogamy relations \cite{zhue}.

In this paper, we study the general monogamy inequalities satisfied by
the concurrence  and concurrence of assistance.
We show that the concurrence of multi-qubit pure states
satisfies some generalized monogamy inequalities.

The concurrence for a bipartite pure state $|\psi\rangle_{AB}$ is given by \cite{s7,s8,af}
\begin{equation}\label{CON}
C(|\psi\rangle_{AB})=\sqrt{2[1-Tr(\rho^2_A)]},
\end{equation}
where $\rho_A$ is the reduced density matrix by tracing over the subsystem $B$,
$\rho_{A}=Tr_{B}(|\psi\rangle_{AB}\langle\psi|)$.
The concurrence is extended to mixed states $\rho=\sum_{i}p_{i}|\psi _{i}\rangle \langle \psi _{i}|$,
$0\leq p_{i}\leq 1$, $\sum_{i}p_{i}=1$, by the convex roof extension,
\begin{equation}\label{CONC}
C(\rho_{AB})=\min_{\{p_i,|\psi_i\rangle\}} \sum_i p_i C(|\psi_i\rangle),
\end{equation}
where the minimum is taken over all possible pure state decompositions of $\rho_{AB}$.

For a tripartite state $|\psi\rangle_{ABC}$, the concurrence of assistance is defined by \cite{ca}
\begin{equation}
C_a(|\psi\rangle_{ABC})\equiv C_a(\rho_{AB})
=\max_{\{p_i,|\psi_i\rangle\}}\sum_ip_iC(|\psi_i\rangle),
\end{equation}
for all possible ensemble realizations of
$\rho_{AB}=Tr_{C}(|\psi\rangle_{ABC}\langle\psi|)=\sum_i p_i |\psi_i\rangle_{AB} \langle \psi_i|$.
When $\rho_{AB}=|\psi\rangle_{AB}\langle \psi|$ is a pure state, then one has
$C(|\psi\rangle_{AB})=C_{a}(\rho_{AB})$.

For an $N$-qubit state $|\psi\rangle_{AB_1...B_{N-1}}$, the concurrence
$C(|\psi\rangle_{A|B_1...B_{N-1}})$ of the state
$|\psi\rangle_{A|B_1...B_{N-1}}$, viewed as a bipartite state
with partition $A$ and $B_1B_2...B_{N-1}$, satisfies the Coffman-Kundu-Wootters
inequality ($N=3$) \cite{CKW} and its generalization to $N-$qubit case.
In \cite{zhue,tjo} it has been shown that the concurrence of a state $\rho$ satisfies a more general monogamy inequality,
$$
C^{\alpha}_{A|B_1B_2...B_{N-1}}\geq C^{\alpha}_{AB_1}+C^{\alpha}_{AB_2}+...+C^{\alpha}_{AB_{N-1}},
$$
where $\rho_{AB_i}=Tr_{B_1...B_{i-1}B_{i+1}...B_{N-1}}(\rho)$,
$C_{A|B_1B_2...B_{N-1}}=C(|\psi\rangle_{A|B_1...B_{N-1}})$, $C_{AB_i}=C(\rho_{AB_i}) (i=1,...,N-1)$ and $\alpha\geq\sqrt{2}$.
The dual inequality in terms of the concurrence of assistance for $N-$qubit states has the form \cite{dualmonogamy},
\begin{equation}\label{ca}
C^2(|\psi\rangle_{A|B_1...B_{N-1}})\leq \sum_{i=1}^{N-1}C^2_a(\rho_{AB_i}).
\end{equation}

The concurrence (\ref{CON}) is related to the linear entropy of a state $\rho$ \cite{esmf},
$$
T(\rho)=1-Tr(\rho^2).
$$
For a bipartite state $\rho_{AB}$, $T(\rho)$ has property \cite{pra74042803},
\begin{equation}\label{T}
|T(\rho_{A})-T(\rho_{B})|\leq T(\rho_{AB})\leq T(\rho_{A})+T(\rho_{B}).
\end{equation}

\begin{theorem}\label{TH1}
For any $2\otimes2\otimes...\otimes 2\otimes2$  pure state $|\psi\rangle=|\psi\rangle_{ABC_1C_2...C_{N-2}}$, we have
\begin{equation}\label{a}
\begin{array}{l}
C^{2}(|\psi\rangle_{AB|C_1C_2...C_{N-2}})\\[2mm]
\displaystyle\geq\max\left\{\sum_{i=1}^{N-2}\left[C^2(\rho_{AC_i})-C_a^2(\rho_{BC_i})\right],\right.\\[5mm]
~~~~~~~~~~~\displaystyle\left.\sum_{i=1}^{N-2}\left[C^2(\rho_{BC_i})-C_a^2(\rho_{AC_i})\right]\right\},
\end{array}
\end{equation}
where $\rho_{AB}=Tr_{C_1...C_{N-2}}(|\psi\rangle\langle\psi|)$,
$\rho_{AC_i}=Tr_{BC_1...C_{i-1}C_{i+1}...C_{N-2}}(|\psi\rangle\langle\psi|)$
and $\rho_{BC_i}=Tr_{AC_1...C_{i-1}C_{i+1}...C_{N-2}}(|\psi\rangle\langle\psi|)$.
\end{theorem}

[Proof]~ For  $2\otimes2\otimes...\otimes2\otimes2$ pure
state $|\psi\rangle_{ABC_1...C_{N-2}}$, one has:
\begin{eqnarray}\label{2}\nonumber
 C^2_{AB|C_1...C_{N-2}}(|\psi\rangle)
 =2(1-Tr{\rho_{AB}})\\[2mm]\nonumber
 \geq |2(1-Tr(\rho_A))- 2(1-Tr(\rho_B))|\\[2mm]\nonumber
 =|C^2_{A|BC_1...C_{N-2}}-C^2_{B|AC_1...C_{N-2}}|,
 \end{eqnarray}
the first inequality is due to the left inequality in (\ref{T}).

For a $2\otimes2\otimes m$ quantum pure state $|\psi\rangle_{ABC}$, it has been shown that
$C^2_{a}(\rho_{AB})= C^2(\rho_{AB})+\tau^C_2(|\psi\rangle_{ABC})$ \cite{022324},
where $\tau^{C}_{2}(|\psi\rangle_{ABC})=C^{2}_{A|BC}-C^{2}_{AB}-C^{2}_{AC}$
is the three tangle of concurrence,
$C_{A|BC}$ is the concurrence of under bipartition $A|BC$ for
pure state $|\psi\rangle_{ABC}$, $\rho_{AB(AC)=Tr_{C(B)}}(|\psi\rangle_{ABC}\langle\psi|)$ and $C_{AB(AC)}=C(\rho_{AB(AC)})$.
Namely,
$$
C^2_{A|BC_1...C_{N-2}}=C^2_{a}(\rho_{AB})+C^2(\rho_{A|C_1...C_{N-2}}).
$$
Hence we have
\begin{eqnarray*}\nonumber
&&C^2_{AB|C_1...C_{N-2}}(|\psi\rangle)\\
&&\geq C^2_{A|BC_1...C_{N-2}}-C^2_{B|AC_1...C_{N-2}}\\
&&\geq C^2(\rho_{A|C_1...C_{N-2}})-\sum_{i=1}^{N-2}C_a^2(\rho_{BC_i})\\
&&\geq\sum_{i=1}^{N-2}C^2(\rho_{AC_i})-\sum_{i=1}^{N-2}C_a^2(\rho_{BC_i}),
\end{eqnarray*}
where the seconde inequality is due to (\ref{ca}).

Similar to above derivation, by using that
$$
C^2_{AB|C_1...C_{N-2}}(|\psi\rangle)\geq C^2_{B|AC_1...C_{N-2}}-C^2_{A|BC_1...C_{N-2}},
$$
we can obtain another inequality in (\ref{a}).
\qed

Theorem 1 shows that the entanglement contained in the pure states
$|\psi\rangle_{ABC_1...C_{N-2}}$ is  related to the sum of entanglement between
bipartitions of the system.

The lower bound in inequalities (\ref{a}) is easily calculable. As an example,
let us consider the four-qubit pure state
\begin{equation}\label{ex}
|\psi\rangle_{ABCD}=\frac{1}{\sqrt{2}}(|0000\rangle+|1001\rangle).
\end{equation}
We have $\rho_{ACD}=Tr_{B}(|\psi\rangle_{ABCD}\langle\psi|)=\frac{1}{2}(|000\rangle+|101\rangle) (\langle000|+\langle101|)$,
$\rho_{BCD}=Tr_{A}(|\psi\rangle_{ABCD}\langle\psi|)=\frac{1}{2}(|000\rangle\langle000|+|001\rangle\langle001|)$,
$C(\rho_{AC})=0$, $C(\rho_{AD})=1$ and $C_a(\rho_{BC})=C_a(\rho_{BD})=0$.
Therefor, $C(|\psi\rangle_{AB|CD})\geq1$, namely,
the state $|\psi\rangle_{ABCD}$ saturates the inequality (\ref{a}).

Theorem 1 gives a monogamy-type lower bound of $C(|\psi\rangle_{AB|C_1C_2...C_{N-2}})$.
According to the subadditivity of the linear entropy, we also have:
\begin{theorem}\label{TH2}
For any $2\otimes2\otimes...\otimes 2\otimes2$  pure state $|\psi\rangle_{ABC_{1}...C_{N-2}}$,
we have
\begin{eqnarray}\label{T2}
&&C^{2}(|\psi\rangle_{AB|C_{1}...C_{N-2}})\\[1mm]\nonumber
&&\leq 2C_a^2(\rho_{AB})+\sum_{i=1}^{N-2}\big(C_a^2(\rho_{AC_i})+C_a^2(\rho_{BC_i})\big),
\end{eqnarray}
where $\rho_{AB}=Tr_{C_1...C_{N-2}}(|\psi\rangle\langle\psi|)$,
$\rho_{AC_i}=Tr_{BC_1...C_{i-1}C_{i+1}...C_{N-2}}(|\psi\rangle\langle\psi|)$
and $\rho_{BC_i}=Tr_{AC_1...C_{i-1}C_{i+1}...C_{N-2}}(|\psi\rangle\langle\psi|)$.
\end{theorem}

[Proof]~ For any $2\otimes2\otimes...\otimes 2\otimes2$  pure state $|\psi\rangle_{ABC_{1}...C_{N-2}}$, one has
\begin{eqnarray}\nonumber
&&C^2_{AB|C_1...C_{N-2}}(|\psi\rangle)\\\nonumber
&&=2(1-Tr(\rho_{AB}))\\\nonumber
&&\leq2(1-Tr(\rho_{A}))+2(1-Tr(\rho_{B}))\\\nonumber
&&=  C^2_{A|BC_1...C_{N-2}}+C^2_{B|AC_1...C_{N-2}}\\\nonumber
&&\leq  2C^2_a(\rho_{AB})+
\sum_{i=1}^{N-2}C_a^2(\rho_{AC_i})+\sum_{i=1}^{N-2}C_a^2(\rho_{BC_i}),
\end{eqnarray}
where the first  inequality is due to the subadditivity of the linear entropy
and the second inequality is due to (\ref{ca}).
\qed

For the four-qubit state (\ref{ex}), we have $C_a(\rho_{AB})=C_a(\rho_{AC})=C_a(\rho_{BC})=C_a(\rho_{BD})=0$ and $C_a(\rho_{AD})=1$.
Then from (\ref{T2}) we have $C(|\psi\rangle_{AB|CD})\leq1$.
The state $|\psi\rangle_{ABCD}$ also saturates the inequality (\ref{T2}).

From the Theorems 1 and 2, in fact, one has
\begin{eqnarray}\label{budengshi}\nonumber
&|&C^2(|\psi\rangle_{A|BC_1...C_{N-2}})-C^2(|\psi\rangle_{B|AC_1...C_{N-2}})|\\
&\leq& C^2(|\psi\rangle_{AB|C_1...C_{N-2}})\\\nonumber
&\leq & C^2(|\psi\rangle_{A|BC_1...C_{N-2}})+C^2(|\psi\rangle_{B|AC_1...C_{N-2}}).
\end{eqnarray}
(\ref{budengshi}) implies that if the systems B (A) and
$C_{1}...C_{N-2}$ are not entangled, then the entanglement between the systems
AB and  $C_{1}...C_{N-2}$ is equal to the entanglement between the systems A
(B) and $C_{1}...C_{N-2}$ for pure states $|\psi\rangle_{ABC_{1}...C_{N-2}}$.

From (\ref{budengshi}), we have
$
C^2(|\psi\rangle_{A|BC_1...C_{N-2}})\leq C^2(|\psi\rangle_{B|AC_1...C_{N-2}})+ C^2(|\psi\rangle_{AB|C_1...C_{N-2}})$
and
$
C^2(|\psi\rangle_{ B|A C_1...C_{N-2}})\leq C^2(|\psi\rangle_{A|BC_1...C_{N-2}})|
+ C^2(|\psi\rangle_{AB|C_1...C_{N-2}}),
$
i.e, the sum of any two of $C^2(|\psi\rangle_{A|BC_1...C_{N-2}})$, $C^2(|\psi\rangle_{B|AC_1...C_{N-2}})$ and $C^2(|\psi\rangle_{AB|C_1...C_{N-2}})$
is greater than or equal to the third.
For convenience, we define $c=C^2_{AB|C_1...C_{N-2}}$,
$a=C^2_{A|BC_1...C_{N-2}}$ and $b=C^2_{B|AC_1...C_{N-2}}$.  In term of (\ref{budengshi}), we can conclude that
for any $2\otimes2\otimes...\otimes 2\otimes2$  pure state $|\psi\rangle_{A|BC_1...C_{N-2}}$,
there are three vectors $\vec{a}$, $\vec{b}$ and $\vec{c}$ with lengthes  $a$, $b$ and $c$,
respectively, which satisfy $\vec{c}= \vec{a}+\vec{b}$.
For example, consider the pure state $|\varphi\rangle_{ABCD}=\frac{1}{\sqrt{3
}}(|0000\rangle+|0010\rangle+|1010\rangle)$. We have $C(|\varphi\rangle_{AB|CD})=\frac{2}{3}$ and
$C(|\varphi\rangle_{A|BCD})=C(|\varphi\rangle_{B|ACD})=\frac{2\sqrt{2}}{3}$.
Then we have $\vec{c}= \vec{a}+\vec{b}$, where $\vec{a}=\frac{2}{9}\vec{e_1}+\frac{2\sqrt{15}}{9}\vec{e_2}$,
$\vec{b}=\frac{2}{9}\vec{e_1}-\frac{2\sqrt{15}}{9}\vec{e_2}$,
$\vec{c}=\frac{4}{9}\vec{e_1}$, $\vec{e_1}$ and $\vec{e_2}$ are two orthogonal basic vectors in a plane.

Now we consider further the generalized monogamy relations in terms of arbitrary partitions
for the $N$-qubit generalized $W$-class states \cite{495301}:
\begin{equation}\nonumber
|W\rangle_{A_1...A_{N}}=a_1 |1\cdots 0\rangle_{A_1...A_{N}} +\cdots +a_N |0\cdots 1\rangle_{A_1...A_{N}},
\end{equation}
where $\sum_{i=1}^{N}|a_i|^2=1$.
One has $C(\rho_{A_pA_q})=C_a(\rho_{A_pA_q})$ $(p\not=q\in\{1,...,N\})$.
Then (\ref{a}) and (\ref{T2}) give rise to
\begin{eqnarray}\label{ws}\nonumber
&&|\sum_{t\in{I}}[C^2(\rho_{A_iA_t})-C^2(\rho_{A_jA_t})]|\\
&&\leq
C^2(|\psi\rangle_{A_iA_j|\overline{A_iA_j}})\\\nonumber
&&\leq 2C^2(\rho_{A_iA_j})+\sum_{t\in{I}}[C^2(\rho_{A_iA_t})+C^2(\rho_{A_jA_t})],
\end{eqnarray}
where $1\leq i<j\leq N$,
$\{\overline{A_iA_j}\}=\{A_1...A_{i-1}A_{i+1}...A_{j-1}A_{j+1}...A_{N}\}$ and $I=\{1, 2,..., i-1, i+1,..., j-1, j+1,..., N\}.$

The inequality Eq.(\ref{ws}) implies that the entanglement (square of concurrence) between $A_iA_j$ and
the other qubits cannot be more than the sum of the individual entanglements between $A_i$ and
each of the $N-1$ remaining qubits and  the the individual entanglements between $A_j$ and each of the $N-1$ remaining qubits.
Take $5$-qubit generalized $W$-class states as an example, one has
$C^2_{AB|C_1C_2C_3}\leq \sum_{t=\{B,C_1,C_2,C_3\}}C^2(\rho_{At})+\sum_{t=\{A,C_1,C_2,C_3\}}C^2(\rho_{Bt}),$
see Fig. 1.
\begin{figure}[htpb]
\centering
\includegraphics[width=8.5cm]{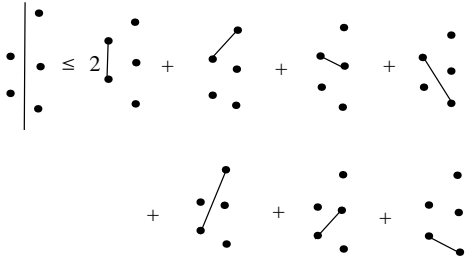}
\caption{{\small  Generalized monogamy for five-qubit generalized $W$-class states.}}
\end{figure}

Now we generalize our results  to the concurrence $C_{ABC_1|C_2...C_{N-2}}$ under partition $ABC_1$ and $C_2...C_{N-2}$ for pure state $|\psi\rangle_{ABC_1C_2...C_{N-2}}$, similar to the theorem 1 and    the theorem 2,
we can obtain  the following corollaries:

{\bf{Corollary 1:}}\label{C1}
For any $N$-qubit  pure state $|\psi\rangle_{ABC_1C_2...C_{N-2}}$, we have
\begin{eqnarray}\label{cor1}
&&C^2(|\psi\rangle_{ABC_1|C_2...C_{N-2}})\\\nonumber
&&\geq\max\big\{\sum_{i=1}^{N-2}C^2(\rho_{AC_i})-\sum_{i=1}^{N-2}C_a^2(\rho_{BC_i}),\\\nonumber
&&~~~\sum_{i=1}^{N-2}C^2(\rho_{BC_i})-\sum_{i=1}^{N-2}C_a^2(\rho_{AC_i})\big\}-\sum_{j\in J}C^2_a(\rho_{C_1j}),
\end{eqnarray}
where $J=\{A, B, C_2,..., C_{N-2}\}$ and $\rho_{C_1j}$ is the reduced density matrix
by tracing over the subsystems except for $C_1$ and $j$.

[proof] For any $N$-qubit  pure state $|\psi\rangle_{ABC_1C_2...C_{N-2}}$, we have
\begin{eqnarray}\nonumber
&C^2&(|\psi\rangle_{ABC_1|C_2...C_{N-2}})\\\nonumber
&=&2(1-Tr(\rho^2_{ABC_1}))\\\nonumber
&\geq& 2(1-Tr(\rho^2_{AB}))-2(1-Tr(\rho^2_{C_1}))\\\nonumber
&=&C^2(|\psi\rangle_{AB|C_1 C_2...C_{N-2}})-C^2(|\psi\rangle_{C_1|ABC_2...C_{N-2}}),
\end{eqnarray}
the first inequality  is due to $T(\rho_{ABC_1})\geq T(\rho_{AB})-T(\rho_{C_1})$.
Combining Theorem 1 and  (\ref{ca}), we obtain (\ref{cor1}).

On the other hand,  from the property of linear entropy, $T(\rho_{ABC_1})\geq T(\rho_{C_1})-T(\rho_{AB}),$
we also can obtain the follow corollary.

{\bf{Corollary 2:}}\label{C2}
For any $N$-qubit  pure state $|\psi\rangle_{ABC_1...C_{N-2}}$, we have
\begin{eqnarray}\label{cor2}
&&C^2(|\psi\rangle_{ABC_1|C_2...C_{N_2}})\\\nonumber
&&\geq C^2{\rho_{AC_1}}+C^2(\rho_{BC_1})
+\sum_{i=2}^{N-2}C^2(\rho_{C_1C_i})\\\nonumber
&&~~~-2C^2_{a}(\rho_{AB})-\sum_{i=1}^{N-2}\left(C^2_{a}(\rho_{AC_i})+C^2_{a}(\rho_{BC_i})\right),\\[2mm]\nonumber
\end{eqnarray}
and
\begin{eqnarray}\label{cor3}
&&C^2(|\psi\rangle_{ABC_1|C_2...C_{N_2}})\\\nonumber
&&\leq 2C_a^2(\rho_{AB})+\sum_{i=1}^{N-2}\left(C_a^2(\rho_{AC_i})+C_a^2(\rho_{BC_i})\right)\\\nonumber
&&~~~+\sum_{j\in J}C_a^2(\rho_{C_1j}),
\end{eqnarray}
where $J$ and $\rho_{C_1j}$ are defined as in Corollary 1.
\medskip

In corollary 2, the upper bound is due to the right inequality of (\ref{ca}) and  $(\ref{T})$.
Analogously, by use of $T(\rho_{ABC_1})\geq |T(\rho_{AC_1})-Tr(\rho_B)|$, $T(\rho_{ABC_1})\geq |T(\rho_{A})-Tr(\rho_{BC_1})|$
and  $T(\rho_{ABC_1})\leq \{T(\rho_{A})+T(\rho_{BC_1}),T(\rho_{AC_1})+T(\rho_B)\}$, one can get more results like (\ref{cor1}), (\ref{cor2})
and (\ref{cor3}). The lower bounds in corollary 1 and corollary 2 are not equivalent.
We consider the following two examples to show that corollary 1 and corollary 2 give rise to different lower bounds.

{\it Example 1:} Let us consider the pure state
$|\psi\rangle_{ABC_1C_2C_3C_4}=\frac{1}{\sqrt{2}}(|000000\rangle+|101000\rangle)$.
We have $C(\rho_{AB})=C(\rho_{AC_i})=C_{a}(\rho_{AC_i})=C(\rho_{C_1C_i})=C_a(\rho_{C_1C_i})=0$ for $i=2,3,4$;
$C(\rho_{BC_i})=C_{a}(\rho_{BC_i})=0$ for $i=1, 2, 3, 4$,
and $C(\rho_{AC_1})=C_a(\rho_{AC_1})=1$. Thus $C(|\psi\rangle)\geq1$ from (\ref{cor1}) and $C(|\psi\rangle)\geq 0$ from (\ref{cor2}).
Namely bound (\ref{cor1}) is better than (\ref{cor2}) in this case.

{\it Example 2:} For the state
$|\psi\rangle_{ABC_1C_2C_3C_4}=\frac{1}{\sqrt{2}}(|000000\rangle+|001100\rangle),$
it is straightforward to calculate that $C(\rho_{AB})=C(\rho_{AC_i})=C(\rho_{BC_i})=C(\rho_{C_1C_3})=C(\rho_{C_1C_4})=C_{a}(\rho_{AC_i})
=C_{a}(\rho_{BC_i})=0$, $i=1, 2, 3, 4$ and $C(\rho_{C_1C_2})=C_a(\rho_{C_1C_2})=1$.
Hence $C(|\psi\rangle)\geq 0$ from (\ref{cor1}), $C(|\psi\rangle)\geq 1$ from (\ref{cor2}) and
bound (\ref{cor2}) is better than (\ref{cor1}) for this state.

We have presented the generalized monogamy relations of concurrence for $N$-qubit systems, by
showing the relations among $C(|\psi\rangle_{AB|C_1...C_{N-2}})$, $C_{AB}$, $C_{AC_i}$, $C_{BC_i}$,
$C_a(\rho_{AC_i})$ and $C_a(\rho_{BC_i})$, $2\leq i\leq N-2$, which give rise to the lower and upper bounds
on the entanglement sharing among the partitions.
Theorem 1 (Theorem 2) gives a lower (an upper) bound of $C(|\psi\rangle_{AB|C_1...C_{N-2}})$. It has been shown that although
$C(|\psi\rangle_{AB|C_1...C_{N-2}})$, $C(|\psi\rangle_{A|B C_1...C_{N-2}})$ and $C(|\psi\rangle_{B|AC_1...C_{N-2}})$ do not satisfy relation like $C^2(|\psi\rangle_{AB|C_1...C_{N-2}})=C^2(|\psi\rangle_{A|B C_1...C_{N-2}}) +C^2(|\psi\rangle_{B|AC_1...C_{N-2}})$, they do satisfy
the triangle inequality: the sum of any two of them is greater or equal to the third.

Entanglement monogamy is a fundamental property of multipartite
entangled states. We have presented a new kind of monogamy relations satisfied by the concurrence of $N$-qubit pure states
under partition $AB$ and $C_1...C_{N-2}$ as well as under partition $ABC_1$ and $C_2...C_{N-2}$.
These relations also give rise to a kind of trade off relations related to the lower and upper bounds of concurrences.
Similar results can be obtained for concurrences under arbitrary partitions $C_{ABC_1...C_i|C_{i+1}...C_{N-2}}$ $(2\leq i< N-1)$.
Such restrictions on entanglement distribution may be also true for other measures of quantum correlations like
negativity or quantum discord.

\bigskip
\noindent{\bf Acknowledgments}\, \,
This work is supported by NSFC under number 11275131. Research Award Fund for
natural science foundation of Shandong province No.ZR2014AP013.

\end{document}